\documentclass[reprint,12pt]{elsarticle}
\usepackage{amsmath}
\usepackage{lineno,hyperref}

\newcommand{\bes}{\begin{subequations}}
\newcommand{\ees}{\end{subequations}}
\newcommand{\bea}{\begin{eqnarray}}
\newcommand{\eea}{\end{eqnarray}}
\newcommand{\ben}{\begin{equation}}
\newcommand{\een}{\end{equation}}
\newcommand{\al}{\alpha}
\newcommand{\ba}{\beta}

\newcommand{\sech}{\mbox{ sech}}
\journal{Physics Letters A}

\begin{document}

\begin{frontmatter}

\title{Nondegenerate soliton solutions in certain coupled nonlinear Schr\"{o}dinger systems}

\author[bdu]{S.~Stalin}
\author[bdu]{R.~Ramakrishnan}
\author[bdu]{M.~Lakshmanan\corref{auth1}}
\address[bdu]{Department of Nonlinear Dynamics, School of Physics, Bharathidasan University, Tiruchirappalli 620024, Tamil Nadu, India}
\cortext[auth1]{Corresponding author; Phone : +91 431 2407093; Fax : +91 431 2407093}
\ead{lakshman.cnld@gmail.com}

\begin{abstract}
In this paper, we report a more general class of nondegenerate soliton solutions, associated with two distinct wave numbers in different modes, for a certain class of physically important integrable two component nonlinear Schr\"{o}dinger type equations through bilinearization procedure. In particular, we consider coupled nonlinear Schr\"{o}dinger (CNLS) equations (both focusing as well as mixed type nonlinearities), coherently coupled nonlinear Schr\"{o}dinger (CCNLS) equations and long-wave-short-wave resonance interaction (LSRI) system. We point out that the obtained general form of soliton solutions exhibit novel  profile structures than the previously known degenerate soliton solutions corresponding to identical wave numbers in both the modes. We show that such degenerate soliton solutions can be recovered from the newly derived 
nondegenerate soliton solutions as  limiting cases. 
\end{abstract}


\end{frontmatter}


\section{Introduction}
 Solitons are localized nonlinear pulses which arise in various nonlinear dispersive media due to the precise balance between nonlinearity and dispersion \cite{1}. Such nonlinear entities remarkably exhibit energy retaining property during  collision process for example in scalar nonlinear Schr\"{o}dinger (NLS) equation where the fundamental soliton  corresponding to intensity is always in a single-hump structure (sech function) characterized by a single wave number \cite{2}. Similar to scalar soliton, the fascinating energy sharing collision exhibiting fundamental multicomponent/vector soliton \cite{j,i} in certain integrable coupled nonlinear Schr\"{o}dinger systems is also described by identical wave numbers in all the modes apart from distinct complex polarization vector constants \cite{j,i}. As a consequence of this, a single-hump structure only occurs in most of the fundamental vector  bright soliton solutions of various CNLS systems.
 
 For instance Manakov type $N$-CNLS equations \cite{i}, mixed $N$-CNLS equations \cite{k}, long-wave-short-wave resonance interaction (LSRI) system  \cite{k01}, etc. are such cases. In contrast to such cases, the coherent coupling among the copropagating optical fields induces a special type of double-hump vector bright soliton in multicomponent CCNLS systems \cite{k1,k11}. In this four wave mixing physical situation also the coherently coupled soliton governed by the same wave number arises in all the modes \cite{k1,k11}. Therefore it is clear that the above mentioned degeneracy in wave numbers always persists in the previously reported vector bright solitons too \cite{j,i,k,k01,k1,k11}.    
 
 Based on the nature of the presence of wave numbers in the multicomponent soliton solution we classify them as degenerate and nondegenerate in the present paper. We call the solitons which propagate in all the modes with identical wave numbers as degenerate vector solitons whereas the solitons with nonidentical wave numbers as nondegenerate vector solitons \cite{ss}. In this context we also note that the terminology nondegenerate solitons has been used in a different context for multi-solitons where the individual constituent solitons travel with distinct velocities in the case of scalar equations such as the Korteweg-deVries, sine-Gordon and NLS equations \cite{ss1}. Then in these cases multi-solitons moving with a single velocity have been referred as degenerate solitons. This is different from our case where we designate solitons with distinct wave numbers in different modes as nondegenerate solitons \cite{ss}. In Refs. \cite{k1} and \cite{k11} one of the present authors and his collaborators have also already discussed these terminologies to classify the coherently coupled solitons as degenerate and nondegenerate based on their intensities:   When the coherently coupled solitons posses the same intensity profile in both the components $q_1$ and $q_2$, they are named as degenerate while the solitons with distinct intensity profiles in the $q_1$ and $q_2$ components are referred as nondegenerate solitons \cite{k1,k11}.  In contrast,, in the present context, the vector solitons already reported in the literature are designated as degenerate class of solitons. In this letter, we intend to show that the above mentioned coupled systems can admit more general class of nondegenerate soliton solutions as in the case of Manakov model reported recently by us \cite{ss} which also finds applications in multicomponent Bose-Einstein condensates \cite{ss2}. Very specifically we derive such new class of soliton solutions for the two component version of CNLS equations, CCNLS equations and LSRI system one by one as we describe below. Their collision property will be reported separately. The procedure we adopt in this work is essentially based on the Hirota's bilinearization method \cite{ss,hir}, while such solutions can also be derived using Darboux transformation method \cite{ss2} or other methods like symmetry based approach \cite{blu}, etc. 
 \section{Nondegenerate bright soliton solutions of CNLS system}
 To start with, we consider the following coupled nonlinear Schr\"{o}dinger equations, 
  \ben
 iq_{j,z}+q_{j,tt}+2\sum_{l=1}^{2}\sigma_l|q_l|^2q_j=0, ~j=1,2,
 \label{1}
 \een
 where $q_j$, $j=1,2$ represent the complex wave amplitudes, with suffices denoting usual partial derivatives. The well known Manakov system \cite{d} arises from Eq. (\ref{1}) when $\sigma_1=\sigma_2=1$, whereas for $\sigma_1=\sigma_2=-1$ and   $\sigma_1=-\sigma_2=1$ turn out to be the  defocusing and mixed type CNLS systems, respectively.  These systems admit bright-bright soliton solutions \cite{j,i}, bright-dark/dark-dark soliton solutions \cite{bd1,k00} and bright-bright/bright-dark/dark-dark soliton solutions \cite{k,bd2,dd,dd1}, respectively, as well as breather and rogue wave type solutions and nonlinear interference patterns \cite{zhao1}. All the above three types of CNLS equations are physically important integrable systems and appear in many physical situations \cite{k,bd1}.      
 
 To derive the nondegenerate bright one-soliton solutions for both the focusing and mixed type CNLS equations as well as to demonstrate the procedure for similar systems, we consider Eq. (\ref{1}) with the following bilinearizing transformations $
 q_{j}(z,t)=\frac{g^{(j)}(z,t)}{f(z,t)}$, $j=1,2$. Here $g^{(j)}$ and $f$ are in general complex and real functions, respectively. Substituting the above transformations in Eq. (1), we obtain the bilinear forms of it as
 \begin{eqnarray}
 D_1~g^{(j)}\cdot f=0, ~j=1,2, ~D_2~ f\cdot f =2\sum_{l=1}^{2}\sigma_lg^{(l)}g^{(j)*}, \label{2}
 \end{eqnarray}
 where $D_1\equiv iD_z+D_t^2$ and $D_2\equiv D_t^2$. The Hirota bilinear operators $D_z$ and $D_t$ are defined as \cite{hir}
 \begin{eqnarray}
 D_z^mD_t^n G\cdot F=\bigg(\frac{\partial}{\partial z}-\frac{\partial}{\partial z'}\bigg)^m\bigg(\frac{\partial}{\partial t}-\frac{\partial}{\partial t'}\bigg)^n G(z,t)\cdot F(z,t)_{|z=z', t=t'} . \label{3}
 \end{eqnarray}
 
 By solving the bilinear equations (\ref{3}) systematically along with the series expansions,
 \begin{eqnarray}
 g^{(j)}=\epsilon g_1^{(j)}+\epsilon^3 g_3^{(j)}+..., ~f=1+\epsilon^2 f_2+\epsilon^4 f_4+..., \label{4}
 \end{eqnarray}
 for the unknown functions $g^{(j)}$ and $f$, we obtain the more general form of nondegenerate soliton solutions for Eq. (\ref{1}) with appropriate nontrivial seed solutions. While constructing the new class of one soliton solution for Eq. (\ref{1}), we find that the above series expansions get truncated as $g^{(j)}=\epsilon g_1^{(j)}+\epsilon^3 g_3^{(j)}$ and $f=1+\epsilon^2 f_2+\epsilon^4 f_4$, by considering the following set of distinct initial seed solutions, $
 g_1^{(1)}=\al_1 e^{\eta_1}$, $g_1^{(2)}=\ba_1 e^{\xi_1}$, $\eta_1=k_1t+ik_1^2z$, $\xi_1=l_1t+il_1^2z$, 
 for the lowest order linear partial differential equations (PDEs), $ig_{1,z}^{(j)}+g_{1,tt}^{(j)}=0$, $j=1,2$. In addition to the latter PDEs we obtain a system of PDEs for the unknown functions $g_3^{(j)}$, $f_2$ and $f_4$, as follows:
 \begin{eqnarray}
 &&\hspace{-0.5cm}O(\epsilon): D_1g_1^{(j)}\cdot 1=0,~O(\epsilon^2):D_2(1\cdot f_2+f_2\cdot 1)=2(\sigma_1 g_1^{(1)}g_1^{(1)*}+\sigma_2 g_1^{(2)}g_1^{(2)*})\nonumber\\
 &&\hspace{-0.5cm}O(\epsilon^3): D_1(g_3^{(j)}\cdot 1+g_1^{(j)}\cdot f_2)=0,~O(\epsilon^5): D_1(g_3^{(j)}\cdot f_2+g_1^{(j)}\cdot f_4)=0,\nonumber\\
 &&\hspace{-0.5cm}O(\epsilon^4):D_2(1\cdot f_4+f_4\cdot 1+f_2\cdot f_2)\nonumber\\&&\hspace{2cm}=2[\sigma_1(g_1^{(1)}g_3^{(1)*}+g_3^{(1)}g_1^{(1)*})+\sigma_2 (g_1^{(2)}g_3^{(2)*}+g_3^{(2)}g_1^{(2)*})]\nonumber\\
 &&\hspace{-0.5cm}O(\epsilon^6):D_2(f_2\cdot f_4+f_4\cdot f_2)=2(\sigma_1 g_3^{(1)}g_3^{(1)*}+\sigma_2 g_3^{(2)}g_3^{(2)*}),\nonumber\\
 &&\hspace{-0.5cm}O(\epsilon^7):D_1g_3^{(j)}\cdot f_4=0,~O(\epsilon^8):D_2f_4\cdot f_4=0,~ j=1,2.\label{5}
 \end{eqnarray}
 The above system of PDEs admits the following solutions:
 \begin{eqnarray}
 &&\hspace{-1cm}g_3^{(1)}=e^{\eta_{1}+\xi_{1}+\xi_{1}^{*}+\Delta_{11}},~ g_3^{(2)}=e^{\xi_{1}+\eta_{1}+\eta_{1}^{*}+\Delta_{12}},~f_2=e^{\eta_{1}+\eta_{1}^{*}+\delta_{1}}+e^{\xi_{1}+\xi_{1}^{*}+\delta_{2}},\nonumber\\
 &&\hspace{-1cm}f_4=e^{\eta_{1}+\eta_{1}^{*}+\xi_{1}+\xi_{1}^{*}+\delta_{11}},~e^{\Delta_{11}}=\frac{\alpha_1|\beta_{1}|^{2}(k_1-l_1)\sigma_{2}}{(k_1+l_1^*)(l_1+l_1^*)^2},~e^{\Delta_{12}}=\frac{\beta_1|\alpha_{1}|^{2}(l_1-k_1)\sigma_{2}}{(k_1^*+l_1)(k_1+k_1^*)^2},\nonumber\\
 &&\hspace{-1cm}e^{\delta_{1}}=\frac{|\alpha_{1}|^{2}\sigma_{1}}{(k_1+k_1^*)^2}, ~
 e^{\delta_{2}}=\frac{|\beta_{1}|^{2}\sigma_{2}}{(l_1+l_1^*)^2},~e^{\delta_{11}}=\frac{|\alpha_{1}|^{2}|\beta_{1}|^{2}|k_1-l_1|^{2}\sigma_{1}\sigma_{2}}{|k_1+l_1^*|^{2}(l_1+l_1^*)^2(k_1+k_1^*)^2}.\label{6}
 \end{eqnarray}
 Note that the other unknown functions in the series expansions (\ref{4}) are found to be zero. Hence the explicit expressions of $g_3^{(j)}$, $f_2$ and $f_4$ constitute the more general form of nondegenerate fundamental soliton solution of CNLS Eq. (\ref{1}) as 
 \begin{eqnarray}
 q_{1}=\frac{\alpha_{1}e^{\eta_{1}}+e^{\eta_{1}+\xi_{1}+\xi_{1}^{*}+\Delta_{11}}}{1+e^{\eta_{1}+\eta_{1}^{*}+\delta_{1}}+e^{\xi_{1}+\xi_{1}^{*}+\delta_{2}}+e^{\eta_{1}+\eta_{1}^{*}+\xi_{1}+\xi_{1}^{*}+\delta_{11}}},\nonumber \\
 q_{2}=\frac{\beta_{1}e^{\xi_{1}}+e^{\xi_{1}+\eta_{1}+\eta_{1}^{*}+\Delta_{12}}}{1+e^{\eta_{1}+\eta_{1}^{*}+\delta_{1}}+e^{\xi_{1}+\xi_{1}^{*}+\delta_{2}}+e^{\eta_{1}+\eta_{1}^{*}+\xi_{1}+\xi_{1}^{*}+\delta_{11}}},\label{7}
 \label{7}
 \end{eqnarray}
 which is exactly of the same form as given for the Manakov equation in \cite{ss}, except that in the various constants $\sigma_1$ and $\sigma_2$ appear explicitly as given in (\ref{6}).
 
 From the above, one can immediately conclude that the obtained solution is nondegenerate because of the fact that  distinct wave numbers $k_1$ and $l_1$ are simultanously present in both the expressions of $q_1$ and $q_2$. The solution (\ref{7}) becomes nondegenerate one bright soliton solution of the Manakov system \cite{ss} if we fix $\sigma_1=\sigma_2=1$ and for the choice $\sigma_1=-\sigma_2=1$, the solution (\ref{7}) is the nondegenerate fundamental soliton solution of the mixed CNLS system. In both the cases the shape of the nondegenerate soliton is described by four nontrivial complex parameters $\al_1$, $\ba_{1}$, $k_1$ and $l_1$. Note that $\alpha_1$ and $\beta_1$ are related to the polarization vectors, $k_{1R}$ and $l_{1R}$ represent the amplitudes while $l_{1I}$ and $k_{1I}$ denote the velocities of the solitons of the two modes $q_1$ and $q_2$, respectively.  
 \begin{figure}[h]
 	\centering
 	\includegraphics[width=0.45\linewidth]{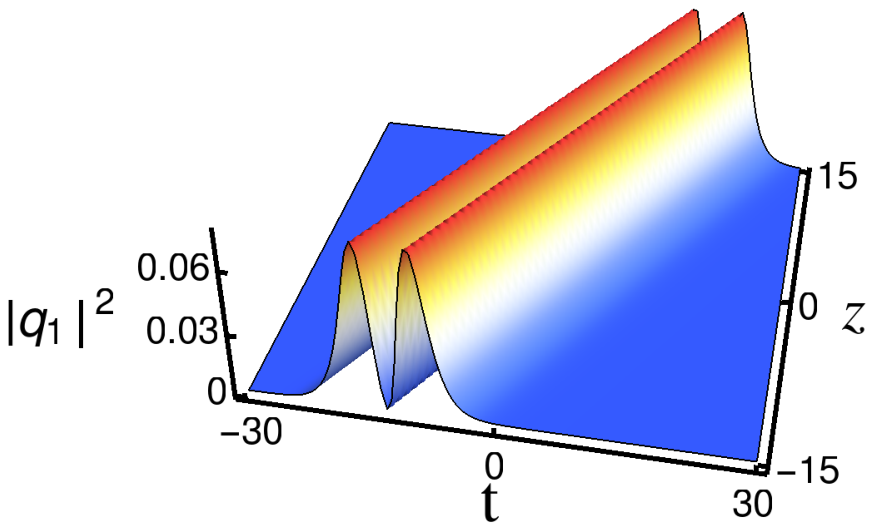}~\includegraphics[width=0.45\linewidth]{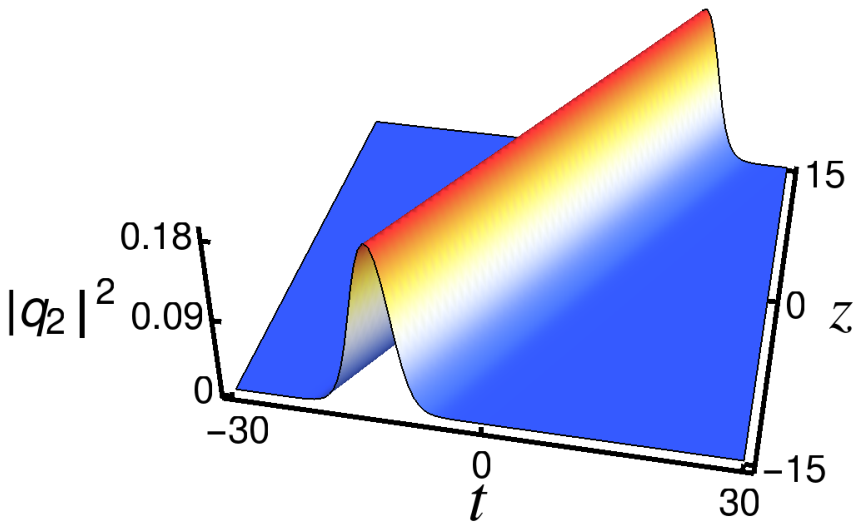}
 	\caption{Nondegenerate symmetric double-hump and single-hump soliton profiles in Manakov system.}
 	\label{fig1}
 \end{figure}

 The distinct wave numbers give rise to two physical situations by restricting the imaginary parts of them. By doing so, we find that the fundamental soliton propagates in the two modes either with identical velocities ($k_{1I}=l_{1I}$) or with non-identical velocities ($k_{1I}\neq l_{1I}$) but with ($k_{1R}\neq l_{1R}$). In the former case the nondegenerate soliton corresponding to the Manakov system admits four distinct nonsingular forms of asymmetric and symmetric profiles which include a single-hump, a double-hump and flattop profiles as we have shown in Ref.\cite{ss}. For the Manakov system, we display  typical double-hump and single-hump profiles in Fig. 1 for the parameter values $k_1=0.333+0.5i$, $l_1=0.55+0.5i$, $\alpha_1=0.5+0.45i$ and $\beta_1=0.5+0.5i$. In contradiction to the Manakov system, the nondegenerate fundamental soliton in the mixed CNLS always shows singular behaviour for arbitrary choice of parameter values, except when $k_1=l_1$. The singularity nature of double-hump soliton profile in this mixed CNLS case is illustrated in Fig. 2 for $k_1=1.2+0.5i$, $l_1=-0.5+0.5i$, $\alpha_1=0.3$ and $\beta_1=i$. The singularity naturally arises  because of the defocusing nonlinearity of the mixed CNLS system.   
    
   \begin{figure}[h]
 	\centering
 	\includegraphics[width=0.45\linewidth]{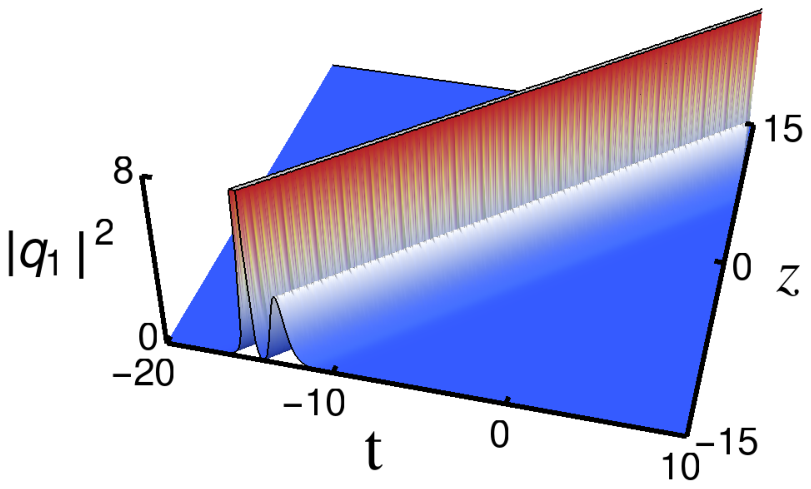}~~\includegraphics[width=0.45\linewidth]{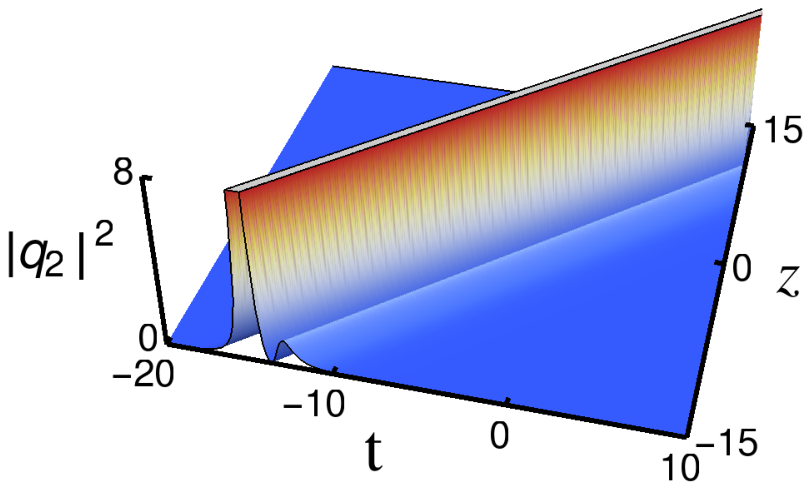}
 \caption{Nondegenerate singular double-hump soliton profiles in mixed CNLS system.}
 	\label{fig2}
 \end{figure}
  \begin{figure}[h]
 	\centering
 	\includegraphics[width=0.45\linewidth]{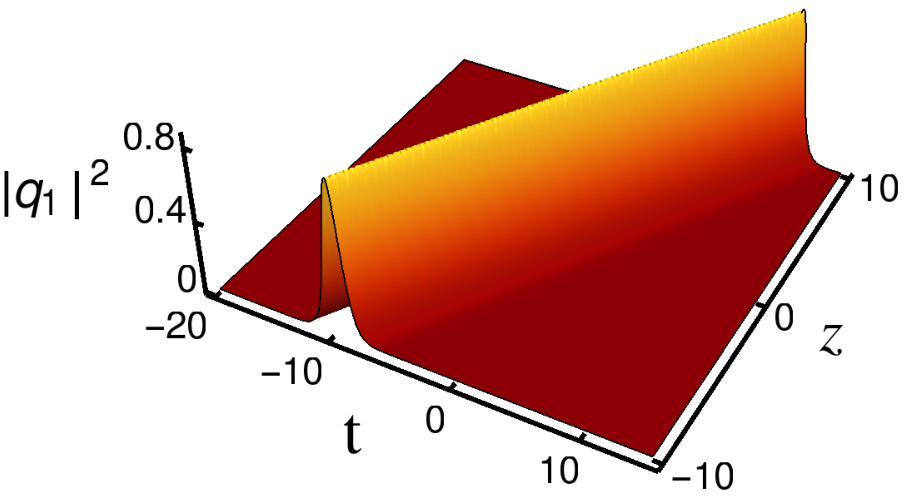}~~\includegraphics[width=0.45\linewidth]{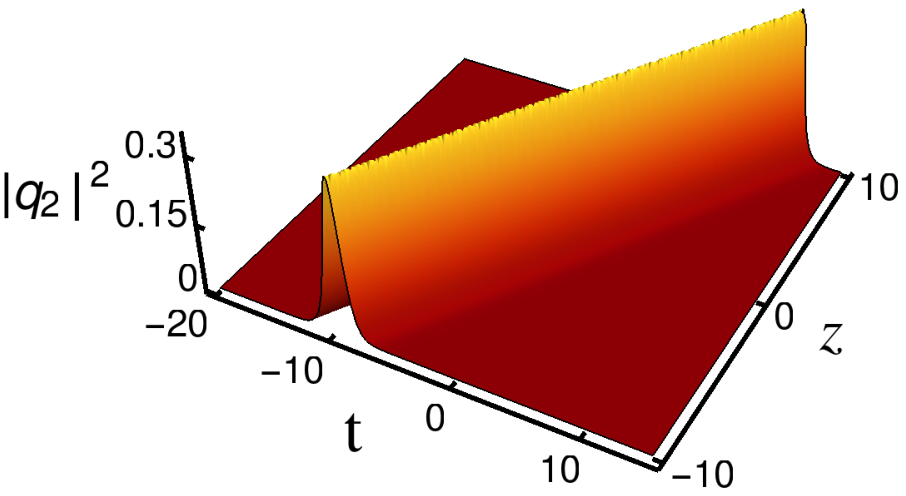}
 	\caption{Degenerate single-hump soliton profiles in Manakov system. }
 	\label{fig3}
 \end{figure}
\begin{figure}[h]
	\centering
	\includegraphics[width=0.45\linewidth]{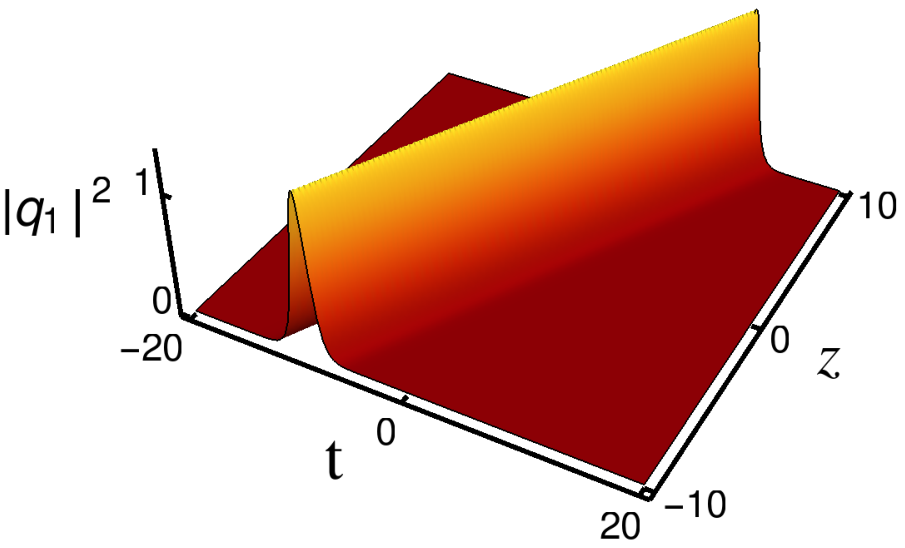}~\includegraphics[width=0.45\linewidth]{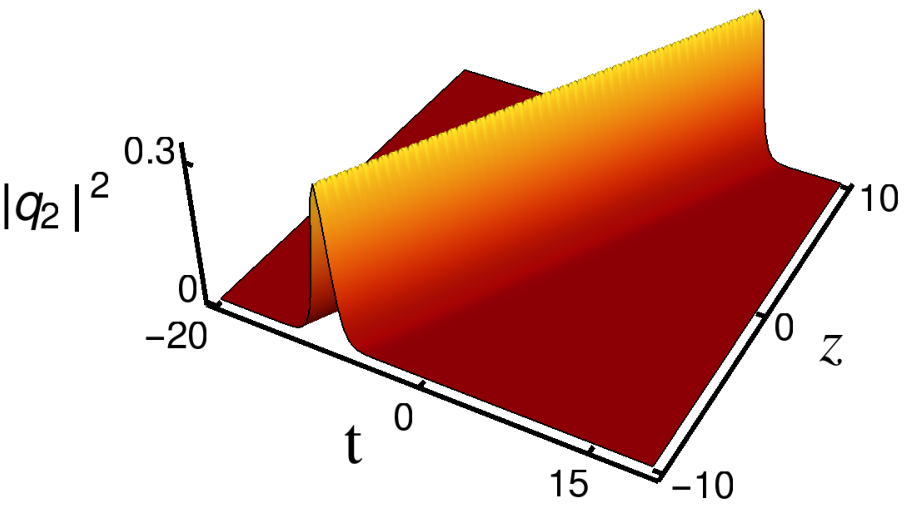}
	\caption{Degenerate non-singular single-hump soliton profiles in mixed CNLS system.}
	\label{fig4}
\end{figure}
 
 If we impose $k_1=l_1$ in Eq. (\ref{7}), the forms of nondegenerate fundamental soliton reduces to the following degenerate bright soliton solution, $q_{j}=\frac{\alpha_{1}^{(j)}e^{\eta_1}}{1+e^{\eta_1+\eta_1^*+R}}\equiv A_jk_{1R}e^{i\eta_{1I}}\sech(\eta_{1R}+\frac{R}{2})$, $j=1,2$ for the Manakov system as well as mixed CNLS system. Here the unit polarization vectors,  $A_1=\frac{\alpha_1}{(\sigma_1|\alpha_1|^2+\sigma_2|\beta_1|^2)^{1/2}}$, $A_2=\frac{\beta_1}{(\sigma_1|\alpha_1|^2+\sigma_2|\beta_1|^2)^{1/2}}$, $\eta_{1R}=k_{1R}(t-2k_{1I}z)$, $\eta_{1I}=k_{1I}t+(k_{1R}^2-k_{1I}^2)z$ and $e^R=\frac{(\sigma_1|\alpha_1|^2+\sigma_2|\beta_1|^2)}{(k_1+k_1^*)^2}$. The amplitude, velocity and the central position of the degenerate fundamental soliton are $A_jk_{1R}$, $2k_{1I}$ and $\frac{R}{2k_{1R}}$, respectively. It is an obvious fact that the degenerate bright soliton solution contains a single complex wave number $k_1$ which allows single-hump profile only. The degenerate fundamental soliton profile of the Manakov system is demonstrated in Fig. 3 for $k_1=1.1+0.5i$, $\alpha_1=1+0.5i$ and $\beta_1=0.5+0.5i$. Similarly for the  mixed CNLS system the non-singular degenerate soliton is shown in Fig. 4 for $k_1=1+0.5i$, $\alpha_1=1$ and $\beta_1=0.5$. As shown in Ref. \cite{k}, the singularity occurs in the degenerate soliton solution of mixed CNLS case when $|\beta_1|>|\alpha_1|$.

 
 \section{Nondegenerate soliton solutions of CCNLS system}
 Now, we consider the following system of two coupled nonlinear Schr\"{o}dinger equations with coherent coupling among the two copropagating fields $q_1$ and $q_2$, 
 \begin{eqnarray}
 iq_{1,z}+q_{1,tt}+\gamma(|q_1|^2+2|q_2|^2)q_1-\gamma q_2^2q_1^*=0, \nonumber\\
 iq_{2,z}+q_{2,tt}+\gamma(2|q_1|^2+|q_2|^2)q_2-\gamma q_1^2q_2^*=0.\label{8}
 \end{eqnarray}
 The terms inside the brackets in Eq. (\ref{8}) correspond to incoherent coupling (self-phase modulation and cross-phase modulation)  between the copropagating fields and the terms $q_2^2q_1^*$ and $q_1^2q_2^*$ correspond to the coherent coupling among the copropagating fields $q_1$ and $q_2$. 
 We note that due to the coherent coupling effect even the degenerate fundamental soliton that is present in the underlying system admits double-hump and flattop profiles apart from the single-hump profile under appropriate parametric choices \cite{k1,k11}. Very interestingly such degenerate coherently coupled soliton undergoes energy switching collision when it interacts with degenerate incoherently coupled soliton \cite{k1,k11}. Equation (\ref{8}) has also been shown to admit breather and rogue wave type solutions too \cite{ccnls1,ccnls2}. Therefore it is interesting to investigate what will happen when the coherently coupled fundamental soliton is characterized by two different wave numbers.  
 
 In order to deduce the appropriate nondegenerate soliton solution to (\ref{8}),  we introduce the bilinear transformation $q_j=\frac{g^{(j)}(z,t)}{f(z,t)}$ with an auxiliary function $s(z,t)$ \cite{k1,k11,k12}.  It results in the following bilinear equations
 \begin{eqnarray}
 D_1 g^{(j)}\cdot f=\gamma g^{(j)*} \cdot s,~D_2 f\cdot f= 2\gamma \sum_{j=1}^{2} |g^{(j)}|^2, s\cdot f =\sum_{j=1}^{2} (g^{(j)})^2,\label{9}
 \end{eqnarray}
 wher $D_1\equiv iD_z+D_t^2$ and $D_2\equiv D_t^2$. We follow the procedure described in \cite{k1,k11} for the  degenerate case but now with the seed solutions $
 g_1^{(1)}=\al_1 e^{\eta_1}$, $g_1^{(2)}=\ba_1 e^{\xi_1}$, $\eta_1=k_1t+ik_1^2z$, $\xi_1=l_1t+il_1^2z$. While doing so, the series expansions get truncated as $g^{(j)}=\epsilon g_1^{(j)}+\epsilon^3 g_3^{(j)}+\epsilon^5 g_5^{(j)}+\epsilon^7 g_7^{(j)}$, $f=1+\epsilon^2 f_2+\epsilon^4 f_4+\epsilon^6 f_6+\epsilon^8 f_8$ and $s=\epsilon^2 s_2+\epsilon^4 s_4+\epsilon^6 s_6$. By substituting the obtained forms of the unknown functions in the truncated series expansions, we get the following general form of nondegenerate coherently coupled fundamental soliton solution of 2-CCNLS system (\ref{8}),
 \begin{eqnarray}
 \hspace{-0.5cm}q_{1}&=&\frac{1}{f}\bigg(\alpha_{1}e^{\eta_{1}}+
 e^{2\eta_{1}+\eta_{1}^{*}+\Delta_{11}}+
 e^{\eta_1^*+2\xi_{1}+\Delta_{12}}+
 e^{\eta_{1}+\xi_{1}+\xi_{1}^{*}+\Delta_{13}}+
 e^{\eta_{1}+2(\eta_{1}^{*}+\xi_{1})+\Delta_{14}}\nonumber\\
 &&~~~+e^{\eta_{1}+2(\xi_{1}+\xi_{1}^{*})+\Delta_{15}}+e^{2\eta_{1}+\eta_{1}^{*}+\xi_{1}+\xi_{1}^{*}+\Delta_{16}}+e^{2(\eta_{1}+\xi_{1}+\xi_{1}^{*})+\eta_{1}^{*}+\Delta_{17}}\bigg),\nonumber\\
 \hspace{-0.5cm}q_2&=&\frac{1}{f}\bigg(\beta_{1}e^{\xi_{1}}+
 e^{2\xi_{1}+\xi_{1}^{*}+\Delta_{21}}+e^{\xi_{1}^{*}+2\eta_{1}+\Delta_{22}}+e^{\xi_{1}+\eta_{1}+\eta_{1}^{*}+\Delta_{23}}+e^{\xi_{1}+2(\xi_{1}^{*}+\eta_{1})+\Delta_{24}}\nonumber\\
 &&~~~+e^{\xi_{1}+2(\eta_{1}^{*}+\eta_{1})+\Delta_{25}}+e^{2\xi_{1}+\xi_{1}^{*}+\eta_{1}+\eta_{1}^{*}+\Delta_{26}}+e^{2(\eta_{1}+\eta_{1}^{*}+\xi_{1})+\xi_{1}^{*}+\Delta_{27}}\bigg),\nonumber
 \end{eqnarray}
  \begin{eqnarray}
 \hspace{-0.5cm}f&=&1+e^{\eta_{1}+\eta_{1}^{*}+\delta_{1}}+
 e^{\xi_{1}+\xi_{1}^{*}+\delta_{2}}+
 e^{2(\eta_{1}+\eta_{1}^{*})+\delta_{3}}+
 e^{2(\eta_{1}+\xi_{1}^{*})+\delta_{4}}+
 e^{2(\xi_{1}+\eta_{1}^{*})+\delta_{5}}\nonumber\\
 &&+e^{2(\xi_{1}+\xi_{1}^{*})+\delta_{6}}+
 e^{(\eta_{1}+\eta_{1}^{*}+\xi_{1}+\xi_{1}^{*})+\delta_{7}}+
 e^{2(\eta_{1}+\eta_{1}^{*})+\xi_{1}+\xi_{1}^{*}+\nu_{1}}\nonumber\\
 &&+e^{2(\xi_{1}+\xi_{1}^{*})+\eta_{1}+\eta_{1}^{*}+\nu_{2}}+
 e^{2(\eta_{1}+\eta_{1}^{*}+\xi_{1}+\xi_{1}^{*})+\nu_{3}}.
 \label{10}
 \end{eqnarray}
 The various constants which appear in the above solution are given by 
 \begin{eqnarray}
&& \hspace{-0.5cm}	e^{\Delta_{11}}=\frac{\gamma \alpha_{1} |\alpha_{1}|^{2}}{2\kappa_{11}}, e^{\Delta_{12}}=\frac{\gamma \alpha_{1}^{*}\beta_{1}^2}{2 \theta_{1}^{*2}}, e^{\Delta_{13}}=\frac{\gamma \alpha_{1} |\beta_{1}|^{2} \rho_{1}}{ \theta_{1}l_{11}}, e^{\Delta_{14}}=\frac{\gamma^{2} \rho_{1}^{2}\alpha_{1}^{*}\beta_{1}^{2}|\alpha_{1}|^{2}}{4 \kappa_{11} \theta_{1}^{*4}},\nonumber\\ &&\hspace{-0.5cm}e^{\Delta_{15}}=\frac{\gamma^{2} \rho_{1}^{2}\alpha_{1} |\beta_{1}|^{4} }{4 l_{11}^{2} \theta_{1}^{2}}, e^{\Delta_{16}}=\frac{\gamma^{2}\rho_{1}^{2}\rho_{1}^{*}\alpha_{1}|\alpha_{1}|^{2}|\beta_{1}|^{2}}{2\kappa_{11}l_{11}\theta_{1}^{2}\theta_{1}^{*}}, e^{\Delta_{17}}=\frac{\gamma^{3}\rho_{1}^{4}{\rho_{1}^{*}}^{2}\alpha_{1}|\alpha_{1}|^{2}|\beta_{1}|^{4}}{8\kappa_{11}l_{11}^{2}\theta_{1}^{4}{\theta_{1}^{*}}^{2}},\nonumber\\
&&\hspace{-0.5cm}e^{\Delta_{21}}=\frac{\gamma \beta_{1} |\beta_{1}|^{2}}{2 l_{11}}, e^{\Delta_{22}}=\frac{\gamma \alpha_{1}^{2} \beta_{1}^{*}}{2\theta_{1}^{2}}, e^{\Delta_{23}}=-\frac{\gamma |\alpha_{1}|^{2} \beta_{1} \rho_{1}}{\theta_{1}^{*} \kappa_{11}}, e^{\Delta_{24}}=\frac{\gamma^{2}\rho_{1}^{2}\alpha_{1}^{2}|\beta_{1}|^{2}\alpha_{1}^{*}}{4l_{11}\theta_{1}^{4}},\nonumber\\
&&\hspace{-0.5cm}e^{\Delta_{25}}=\frac{\gamma^{2}\rho_{1}^{2}|\alpha_{1}|^{4}\beta_{1}}{4\kappa_{11}^{2}\theta_{1}^{*2}}, e^{\Delta_{26}}=-\frac{\gamma^{2}\rho_{1}^{2}\rho_{1}^{*}\beta_{1}|\alpha_{1}|^{2}|\beta_{1}|^{2}}{2\kappa_{11}l_{11}\theta_{1}\theta_{1}^{*2}}, e^{\Delta_{27}}=\frac{\gamma^{3}\rho_{1}^{4}\rho_{1}^{*2}\beta_{1}|\alpha_{1}|^{4}|\beta_{1}|^{2}}{8\kappa_{11}^{2}l_{11}\theta_{1}^{2}\theta_{1}^{*4}},\nonumber\\
&&\hspace{-0.5cm} e^{\delta_{1}}=\frac{\gamma |\alpha_{1}|^{2}}{\kappa_{11}}, e^{\delta_{2}}=\frac{\gamma |\beta_{1}|^{2}}{l_{11}}, e^{\delta_{3}}=\frac{\gamma^{2} |\alpha_{1}|^{4}}{4 \kappa_{11}^{2}}, e^{\delta_{4}}=\frac{\gamma^{2}\alpha_{1}^{2} \beta_{1}^{*2}}{4 \theta_{1}^{4}}, e^{\delta_{5}}=\frac{\gamma^{2} \alpha_{1}^{*2} \beta_{1}^{2}}{4\theta_{1}^{*4}},\nonumber\\
&&\hspace{-0.5cm}e^{\delta_{6}}=\frac{\gamma^{2} |\beta_{1}|^{4}}{4 l_{11}^{2}}, e^{\delta_{7}}=\frac{\gamma^{2} |\rho_{1}|^{2} |\alpha_{1}|^{2} |\beta_{1}|^{2}}{\kappa_{11} l_{11} |\theta_{1}|^{2}}, e^{\nu_{1}}=\frac{\gamma^{3}|\rho_{1}|^{4}|\alpha_{1}|^{4}|\beta_{1}|^{2}}{4\kappa_{11}^{2}l_{11}|\theta_{1}|^{4}},\nonumber\\
&&\hspace{-0.5cm}e^{\nu_{2}}=\frac{\gamma^{3}|\rho_{1}|^{4}|\alpha_{1}|^{2}|\beta_{1}|^{4}}{4\kappa_{11}l_{11}^{2}|\theta_{1}|^{2}},~ e^{\nu_{3}}=\frac{\gamma^{4}|\rho_{1}|^{8}|\alpha_{1}|^{4}|\beta_{1}|^{4}}{16\kappa_{11}^{2}l_{11}^{2}|\theta_{1}|^{8}},   l_{11}=(l_1+l_1^*)^2,
 \nonumber\\
 &&\hspace{-0.5cm}\theta_{1}=(k_1+l_1^*),~\rho_{1}=(k_1-l_1),~\kappa_{11}=(k_1+k_1^*)^2.\nonumber
 \end{eqnarray} 
  
 The auxiliary function is obtained as
 $s=\alpha_1^2e^{2\eta_1}+\beta_1^2e^{2\xi_1}+e^{2\eta_1+\xi_1+\xi_1^*+\phi_1}+e^{2\xi_1+\eta_1+\eta_1^*+\phi_2}+e^{2(\eta_1+\eta_1^*+\xi_1)+\phi_3}+e^{2(\eta_1+\xi_1^*+\xi_1)+\phi_4}$, $e^{\phi_1}=\frac{\gamma\rho_{1}^2\alpha_1^2|\beta_1|^2}{\theta_{1}^2l_{11}}$, $e^{\phi_2}=\frac{\gamma\rho_{1}^2\beta_1^2|\alpha_1|^2}{\theta_{1}^{*2}\kappa_{11}}$, $e^{\phi_3}=\frac{\gamma^2\rho_{1}^4\beta_1^2|\alpha_1|^4}{4\theta_{1}^{*4}\kappa_{11}^2}$, $e^{\phi_4}=\frac{\gamma^2\rho_{1}^4\alpha_1^2|\beta_1|^4}{4\theta_{1}^{4}l_{11}^2}$. The  already reported degenerate coherently coupled fundamental one-soliton solution \cite{k1,k11} of Eq. (\ref{8}) is obtained by restricting $k_1=l_1$ in Eq. (\ref{10}). This leads to 
 $q_1=\frac{\alpha_1e^{\eta_1}+e^{2\eta_1+\eta_1^*+\Delta_{11}}}{1+e^{\eta_1+\eta_1^*+\delta_1}+e^{2(\eta_1+\eta_1^*)+\delta_2}}$, $q_2=\frac{\beta_1e^{\eta_1}+e^{2\eta_1+\eta_1^*+\Delta_{12}}}{1+e^{\eta_1+\eta_1^*+\delta_1}+e^{2(\eta_1+\eta_1^*)+\delta_2}}$, $e^{\Delta_{11}}=\frac{\gamma \alpha_1^*(\alpha_1^2+\beta_1^2)}{2\kappa_{11}}$, $e^{\Delta_{12}}=\frac{\gamma \beta_1^*(\alpha_1^2+\beta_1^2)}{2\kappa_{11}}$, $e^{\delta_1}=\frac{\gamma(|\alpha_1|^2+|\beta_1|^2)}{\kappa_{11}}$, $e^{\delta_2}=\frac{\gamma^2|\alpha_1^2+\beta_1^2|^2}{4\kappa_{11}^2}$. The auxiliary function is reduced as $s=(\alpha_1^2+\beta_1^2)e^{2\eta_1}$. 
 \begin{figure}[h]
 	\centering
 	\includegraphics[width=0.45\linewidth]{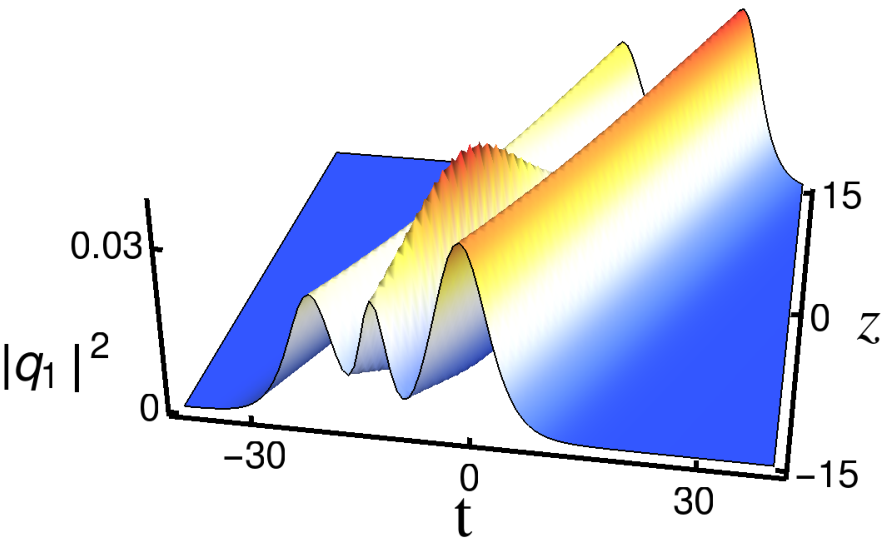}~~\includegraphics[width=0.45\linewidth]{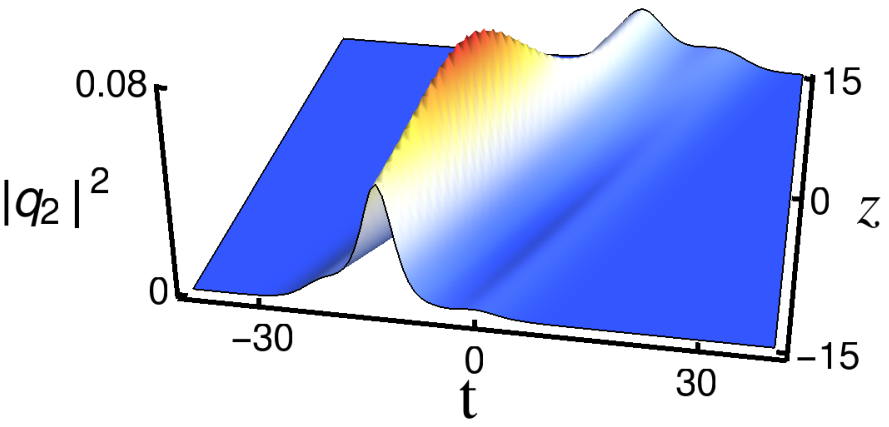}
 	\caption{Breathing type triple-hump profile of nondegenerate soliton in the CCNLS system.}
 	\label{fig5}
 \end{figure}

  From the solution (\ref{10}), it is easy to identify that the shape of the nondegenerate coherently coupled fundamental soliton (\ref{10}) is also governed by two arbitrary complex parameters $\alpha_1$ and $\ba_1$ and two distinct complex wave numbers $k_1$ and $l_1$. The solution (\ref{10}) admits various novel profiles , such as a quadruple-hump, a triple-hump, a double-hump and a single-hump profiles under appropriate restrictions on the wave parameters. This is due to the presence of additional wave number and the four wave mixing effect.  As an example, we display a nontrivial breathing type triple-hump shaped soliton profiles in Fig. 5 for the parameters  $\gamma=2$, $k_1=0.21+0.5i$, $l_1=0.29+0.5i$, $\alpha_1=0.95+0.5i$ and $\beta_1=0.97-i$. By tuning the relative separation distance it is also possible to separate a single-hump and a double-hump from this triple-hump profile. However, a distinct double-hump profile only occurs in the  degenerate case. This is due to the presence of a single wave number apart from  two arbitrary constants $\al_1$ and $\beta_{1}$. A typical degenerate flattop soliton in $q_1$ component and a double-hump profile in $q_2$ component is illustrated in Fig. 6 for $\gamma=2$, $k_1=l_1=0.5+0.5i$, $\alpha_1=0.72+0.5i$ and $\beta_1=0.5-0.42i$.
 \begin{figure}[h]
 	\centering
 	\includegraphics[width=0.45\linewidth]{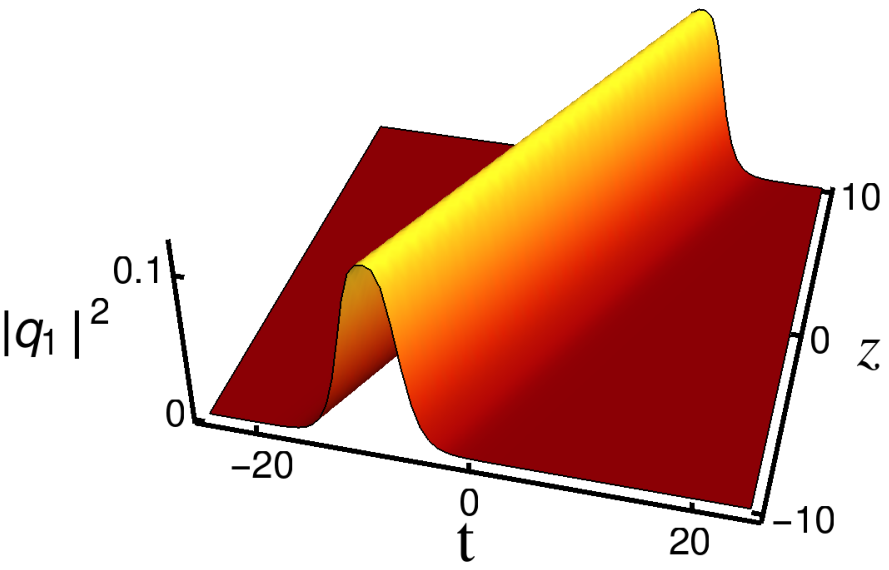}~~\includegraphics[width=0.45\linewidth]{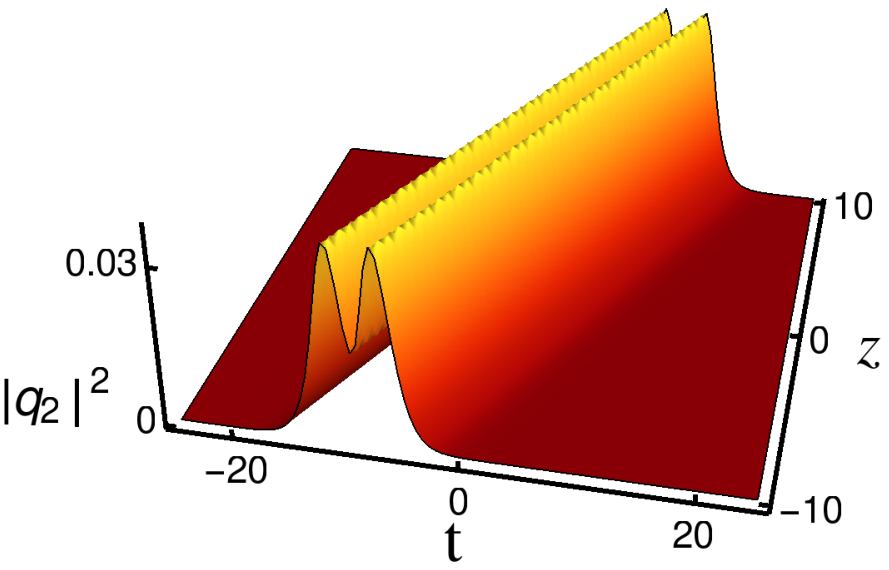}
 	\caption{Flattop-double-hump profiles of degenerate solitons in the CCNLS system. }
 	\label{fig6}
 \end{figure}

  \section{Nondegenerate soliton solution of LSRI system}
  Finally we intend to derive the nondegenerate fundamental soliton solution for the following long-wave short-wave resonance interaction system, namely the 2-component Yajima-Oikawa system \cite{k13} with general form of nonlinearity, 
 \begin{eqnarray}
 \hspace{-0.5cm} iS_t^{(1)}+S_{xx}^{(1)}+LS^{(1)}=0,~iS_t^{(2)}+S_{xx}^{(2)}+LS^{(2)}=0,~L_t=\sum_{l=1}^{2}\sigma_l (|S^{(l)}|^2)_x.\label{11}
 \end{eqnarray}   
 In the above, $S^{(l)}$'s, $l=1,2$, are short-wave components and $L$ is the long-wave component and suffices denote partial derivatives, while $\sigma_l$'s are arbitrary real parameters. Further $\sigma_l=+1$, $\sigma_l=-1$, $l=1,2$, and $\sigma_1=-\sigma_2=1$ correspond to positive, negative and mixed positive-negative nonlinearities. Both nondegenerate and degenerate solitons arise in the present short-wave components also due to the balance between their dispersion and nonlinear interactions of the short-waves with a long-wave. In contrast to the previous case, the formation of nondegenerate and degenerate solitons arises in the long-wave component due to the interaction of the short-wave components. In the present 2-component LSRI system also the solitons in the short-wave components as well as long-wave component are degenerate characterized by a single wave number. To overcome this degeneracy we take the modified form of seed solutions, involving two distinct wave numbers, in the nondegenerate soliton solution construction process. We note that the above LSRI system admits rogue wave solutions also\cite{lsri1}.      
 
 To construct the nondegenerate one-soliton solution we again  bilinearize Eq. (\ref{11}) through  the following  transformations, $ S^{(l)}(x,t)=\frac{g^{(l)}(x,t)}{f(x,t)}$, $l=1,2$, $L=2\frac{\partial^2}{\partial x^2}\ln f(x,t)$. We obtain  the following bilinear forms: 
 \bea
 D_1 g^{(l)}\cdot f=0, l=1,2, ~D_2 f\cdot f=\sum_{n=1}^{2}\sigma_n |g^{(n)}|^2,\label{12}
 \eea
 where $D_1\equiv iD_t+D_x^2$ and $D_2\equiv D_xD_t$. With the modified forms of seed solutions $
 g_1^{(1)}=\al_1 e^{\eta_1}$, $g_1^{(2)}=\ba_1 e^{\xi_1}$, $\eta_1=k_1x+ik_1^2t$, $\xi_1=l_1x+il_1^2t$, we find that the series expansions which are given in \cite{k01} get terminated as $g^{(l)}=\epsilon  g^{(l)}_1+\epsilon^3  g^{(l)}_3$, $f=1+\epsilon^2 f_2+\epsilon^4 f_4$.
 The explicit forms of the unknown functions lead to the following nondegenerate fundamental soliton solution, 
 \begin{eqnarray}
 &&S^{(1)}=\frac{g_1^{(1)}+g_3^{(1)}}{1+f_2+f_4}=\frac{\alpha_{1}e^{\eta_{1}}+e^{\eta_{1}+\xi_{1}+\xi_{1}^{*}+\mu_{11}}}
 {1+e^{\eta_{1}+\eta_{1}^{*}+R_{1}}+e^{\xi_{1}+\xi_{1}^{*}+R_{2}}+e^{\eta_{1}+\eta_{1}^{*}+\xi_{1}+\xi_{1}^{*}+R_{3}}},\nonumber\\
 &&S^{(2)}=\frac{g_1^{(2)}+g_3^{(2)}}{1+f_2+f_4}=\frac{\beta_{1}e^{\xi_{1}}+e^{\xi_{1}+\eta_{1}+\eta_{1}^{*}+\mu_{12}}}
 {1+e^{\eta_{1}+\eta_{1}^{*}+R_{1}}+e^{\xi_{1}+\xi_{1}^{*}+R_{2}}+e^{\eta_{1}+\eta_{1}^{*}+\xi_{1}+\xi_{1}^{*}+R_{3}}},\nonumber\\
&&L=\frac{2}{f^2}\bigg((k_1+k_1^*)^2e^{\eta_{1}+\eta_{1}^{*}+R_{1}}+(l_1+l_1^*)^2e^{\xi_{1}+\xi_{1}^{*}+R_{2}}+e^{\eta_{1}+\eta_{1}^{*}+\xi_{1}+\xi_{1}^{*}+R_{4}}\nonumber\\
 &&~~~~~~~~~+e^{2(\eta_{1}+\eta_{1}^{*})+\xi_{1}+\xi_{1}^{*}+R_1+R_{3}}+e^{\eta_{1}+\eta_{1}^{*}+2(\xi_{1}+\xi_{1}^{*})+R_2+R_{3}}\bigg),\nonumber\\
 &&f={(1+e^{\eta_{1}+\eta_{1}^{*}+R_{1}}+e^{\xi_{1}+\xi_{1}^{*}+R_{2}}+e^{\eta_{1}+\eta_{1}^{*}+\xi_{1}+\xi_{1}^{*}+R_{3}})},	\label{13}
 \end{eqnarray} 
 where  $ e^{\mu_{11}}=\frac{i\al_1|\ba_1|^2\sigma_2(l_1-k_1)}{2(k_1+l_1^*)(l_1-l_1^*)(l_1+l_1^*)^2}$, ~ $e^{\mu_{12}}=\frac{i\ba_1|\al_1|^2\sigma_1(k_1-l_1)}{2(k_1^*+l_1)(k_1-k_1^*)(k_1+k_1^*)^2}$,~
 $e^{R_1}=\frac{|\al_1|^2\sigma_1}{2i(k_1+k_1^*)^2(k_1-k_1^*)}$, $e^{R_2}=\frac{|\ba_1|^2\sigma_2}{2i(l_1+l_1^*)^2(l_1-l_1^*)}$, $e^{R_3}=-\frac{|\al_1|^2|\ba_1|^2|k_1-l_1|^2\sigma_1\sigma_2}{4|k_1+l_1^*|^2(k_1-k_1^*)(l_1-l_1^*)(k_1+k_1^*)^2(l_1+l_1^*)^2}$, $e^{R_4}=-2(k_1+k_1^*)(l_1+l_1^*)(e^{R_1+R_2}-e^{R_3})+((k_1+k_1^*)^2+(l_1+l_1^*)^2)(e^{R_1+R_2}+e^{R_3})$.
 
 The nondegenerate fundamental soliton in the 2-component LSRI sytem is also governed by four non-trivial arbitrary complex parameters $\alpha_1$, $\ba_1$, $k_1$ and $l_1$.  The amplitudes of the nondegenerate fundamental solitons in the short-wave components are $4k_{1R}A_1\sqrt{k_{1I}}$,  $4l_{1R}A_2\sqrt{l_{1I}}$. Here $A_1=\frac{-i\sqrt{\al_1}}{\sqrt{\sigma_1\al_1^*}}$, $A_2=\frac{-i\sqrt{\beta_1}}{\sqrt{\sigma_2\ba_1^*}}$ are unit polarization vectors of the two short-wave components. In the present case the velocity of the nondegenerate fundamental soliton is characterized by  the imaginary  parts of the wave numbers $k_1$ and $l_1$. Very interestingly in the present LSRI system, the nondegenerate fundamental soliton exhibits amplitude dependent velocity property like the KdV-soliton. The degenerate soliton also possesses this unusual property \cite{k01}. As a consequence of this property the taller  nondegenerate soliton will propagate faster than the shorter one.  To get the regular solution the quantities $e^{R_1}$, $e^{R_2}$ and $e^{R_3}$ in (\ref{13}) should be positive. To achive this, we fix $k_{1I},l_{1I}<0$, $k_{1I},l_{1I}>0$ and $k_{1I}<0$, $l_{1I}>0$ for the positive ($\sigma_l>0$), negative ($\sigma_l<0$) and mixed type ($\sigma_1=1$, $\sigma_2=-1$) nonlinearities, respectively. In all the three cases, we observe that the  nondegenerate fundamental soliton in the present system admits double-hump profiles similar to  nondegenerate soliton of Manakov system. We depict asymmetric double-hump profiles of nondegenerate one-soliton in Fig. 7 for the parameters $k_1=0.3-0.5i$, $l_1=0.35-0.5i$, $\alpha_1=0.8$, $\beta_1=0.5$ and $\sigma_1=\sigma_2=1$.    
 \begin{figure}[h]
 	\centering
 	\includegraphics[width=0.33\linewidth]{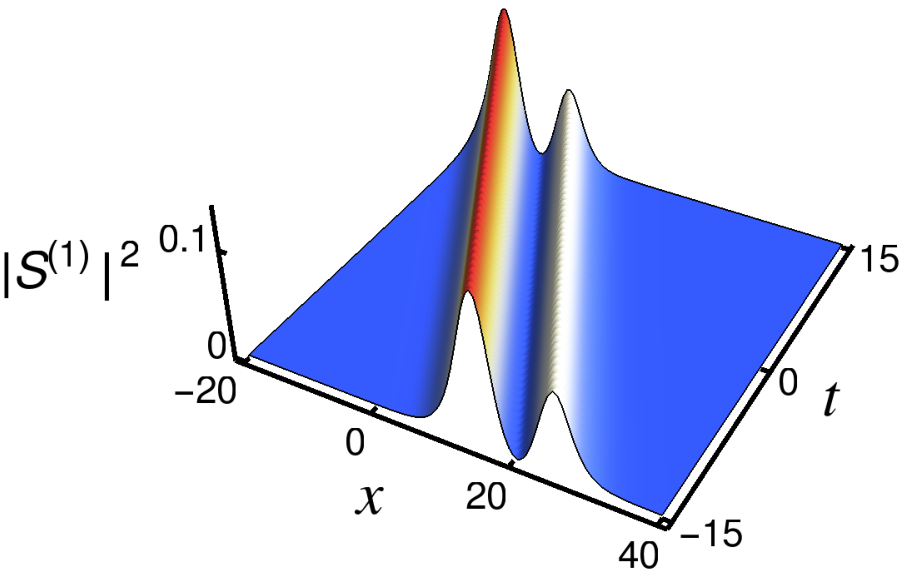}~~\includegraphics[width=0.33\linewidth]{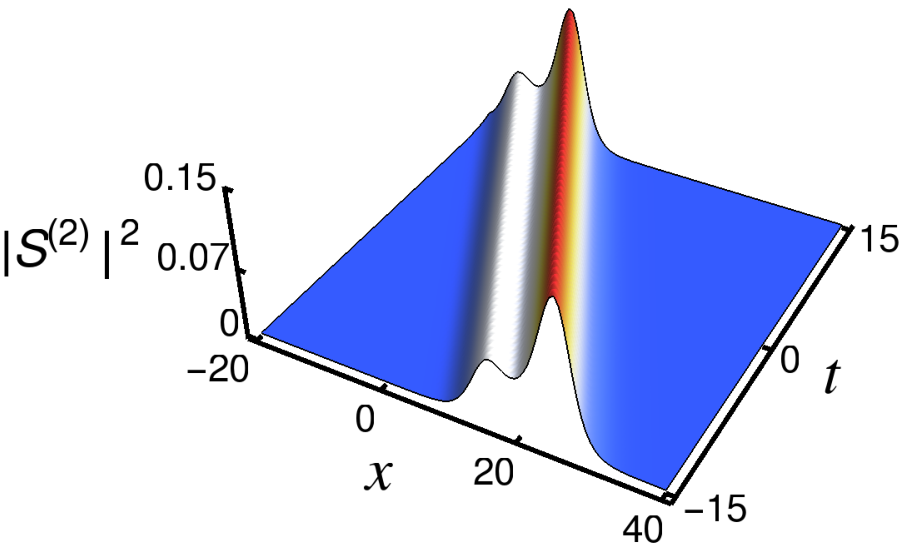}\includegraphics[width=0.33\linewidth]{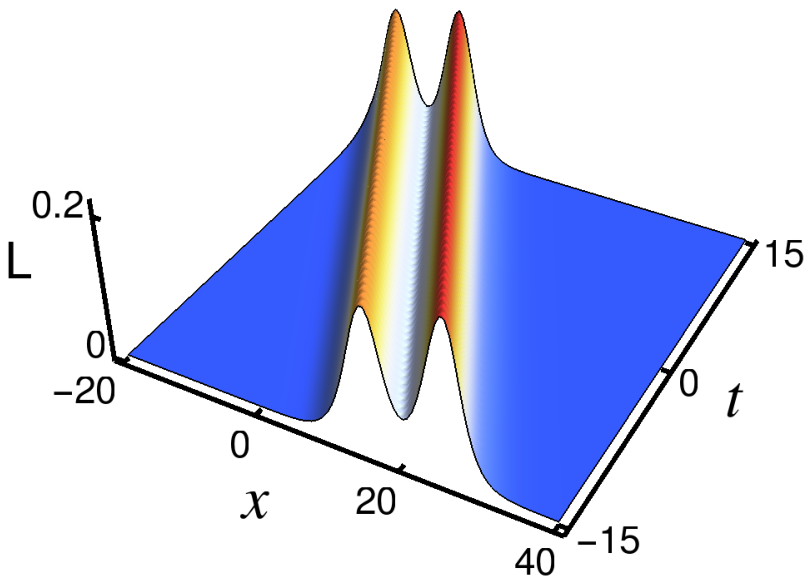}
 	\caption{Nondegenerate asymmetric  double-hump soliton profiles in the two short-wave components and the long-wave component.}
 	\label{fig7}
 \end{figure} 
 \begin{figure}[h]
 	\centering
 	\includegraphics[width=0.35\linewidth]{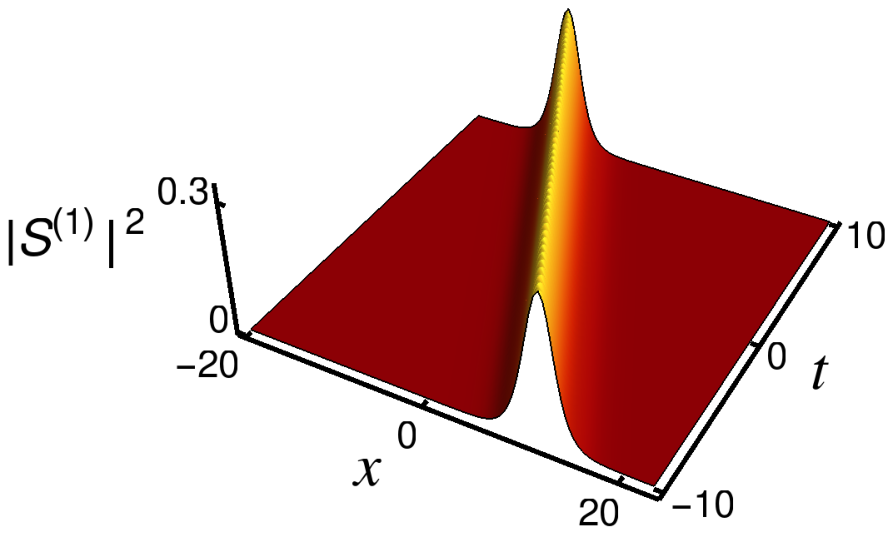}~\includegraphics[width=0.35\linewidth]{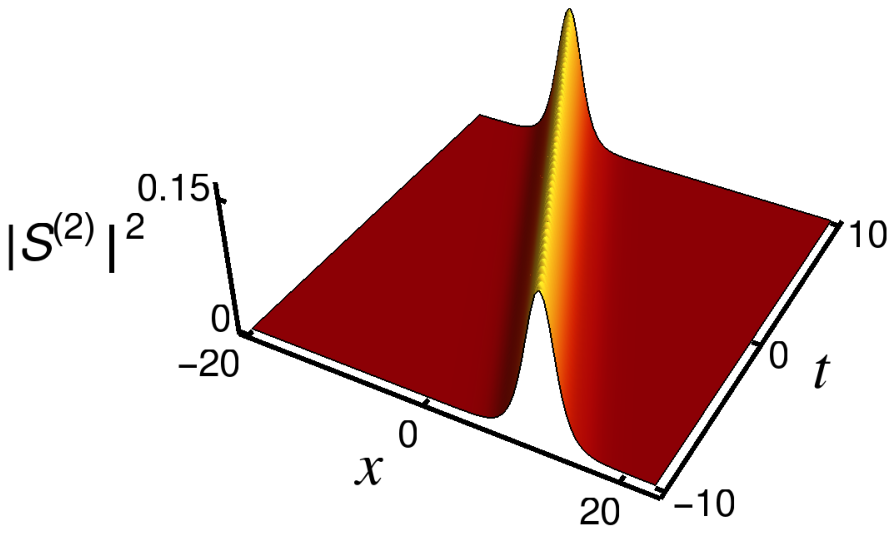}~\includegraphics[width=0.33\linewidth]{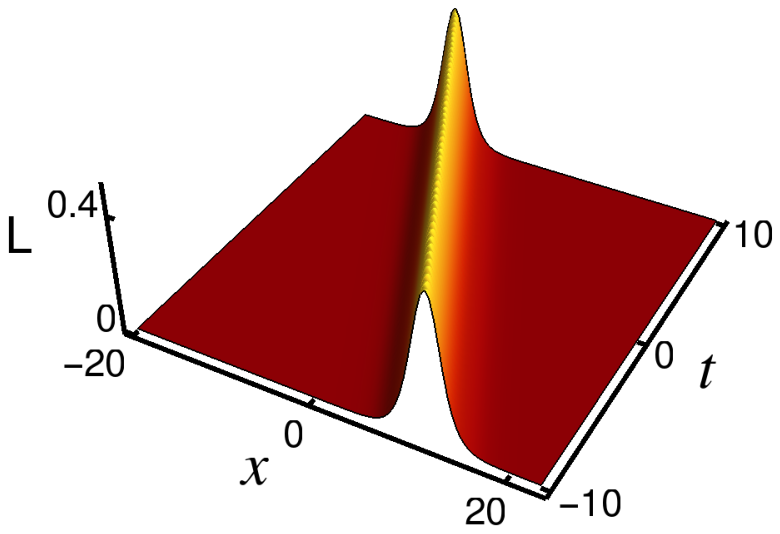}
 	\caption{Degenerate single-hump soliton profiles in both the short-wave components and the long-wave component. }
 	\label{fig8}
 \end{figure} 

 We recover degenerate soliton solution of Eq. (\ref{11}) by substituting the limit $k_1=l_1$ in Eq. (\ref{13}). This results in the following degenerate fundamental soliton forms: $S^{(l)}=2A_lk_{1R}\sqrt{k_{1I}}e^{i(\eta_{1I}+\frac{\pi}{2})}\sech(\eta_{1R}+\frac{R}{2})$, $L=2k_{1R}^2\sech^2(\eta_{1R}+\frac{R}{2})$, $l=1,2$. Here $A_1=\frac{\alpha_1}{(\sigma_1|\alpha_1|^2+\sigma_2|\beta_1|^2)^{1/2}}$, $A_2=\frac{\beta_1}{(\sigma_1|\alpha_1|^2+\sigma_2|\beta_1|^2)^{1/2}}$, $\eta_{1R}=k_{1R}(t+2k_{1I}z)$, $\eta_{1I}=k_{1I}t+(k_{1R}^2-k_{1I}^2)z$, $e^R=\frac{-(\sigma_1|\alpha_1|^2+\sigma_2|\beta_1|^2)}{16k_{1R}^2k_{1I}}$.  As discussed in \cite{k01}, the degenerate soliton in both the short-wave components and the long-wave component admits only a single-hump profile. A typical graph of such single-hump profile is shown in Fig. 8 for $k_1=0.5-0.5i$, $\alpha_1=0.5$, $\beta_1=0.35$ and $\sigma_1=\sigma_2=1$.
 \section{Conclusion}
 In this work, we have thus derived more general forms of nondegenerate fundamental bright solitons corresponding to non-identical wave-numbers for certain physically important integrable coupled systems. In particular we have considered the two component version of the Manakov system, mixed CNLS system, coherently coupled NLS system and long-wave short-wave resonance interaction system. We find that the obtained nondegenerate bright soliton solution admits various novel structures compared to the corresponding degenerate counterparts. The interesting collision dynamics of such nondegenerate solitons will be presented elsewhere.  
\section*{Acknowledgements}
The works of SS and ML are supported by the DST-SERB research project (EMR/2014/001076), Government of India. RR and ML acknowledge the financial support under a DST-SERB Distinguished Fellowship program (SB/DF/04/2017) to ML.
\section*{References}

\bibliography{mybibfile}

\end{document}